\documentstyle[12pt]{article}
\newcommand\be{\begin{equation}}
\newcommand\ee{\end{equation}}
\newcommand\bey{\begin{eqnarray}}
\newcommand\eey{\end{eqnarray}}
\begin{document}
\setlength{\baselineskip}{12pt}
\title{\hfill{\rm\normalsize{HD--TVP--94--7}}}
\title{
Effective light front quantization of scalar field theories and two-dimensional
electrodynamics}
\author{E.V. Prokhvatilov\thanks{Permanent address: Department of Theoretical
Physics, Institute of Physics, St. Petersburg University, Ulyanovskaya 1, St.
Petersburg 198904, Russia}, H.W.L. Naus\thanks{Supported in part by the Federal
Ministry of Research and Technology (BMFT) under contract number 06 HD 729}
and  H.--J. Pirner
\\\\
Institut f\"ur Theoretische Physik\\
Universit\"at Heidelberg}

\date{ }

\vspace{1.0cm}

\maketitle

\begin{abstract}
\setlength{\baselineskip}{18pt}
\noindent
{We introduce  a new method to include condensates in the light-cone
Hamiltonian.
By using a Gaussian approximation to the ordinary vacuum in a theory close to
the
light front, we derive an effective Hamiltonian on the light cone, which has
new terms reflecting the nontriviality of the vacuum. We demonstrate our method
for scalar $\phi^4$-theory and the massive Schwinger model. }
\end{abstract}
\newpage
\section{Introduction}

The idea of quantizing field theories on the light front (i.e. on the
hyperplane
tangent to the light-cone) was put forward by Dirac~\cite{dir}. He pointed out
that in
such a formulation, the part of the Lorentz symmetry described kinematically is
maximal. In other words, the number of generators of the Poincar\'e group,
which
depend on the dynamics, is minimal. Instead of Lorentz coordinates
$x^\mu(\mu=0,1,2,3)$ Dirac used the light-like coordinates:
\bey\label{1}
x^\pm&=&\frac{1}{\sqrt2}(x^0\pm x^3),\nonumber\\
x^\bot&=& x^{1,2}.
\eey
The coordinate $x^+$ plays the role of the time. The
subgroup of the Poincar\'e group consisting of the generators
$M_{12},M_{+-},M_{-\bot}$
and $P^+,P_\bot$ is dynamically independent.
This maximal amount of kinematical symmetry is related to the
trivial structure of the vacuum in this formulation~\cite{kla}. Indeed the
vacuum is
identified  with the lowest eigenstate of the momentum $P^+\geq 0$. The Fock
space
constructed over this vacuum~\cite{kla} can be used to solve the eigenvalue
problem for
the mass (squared) operator: $m^2=2P^+P^--P^2_\bot$. For states with fixed
$P^+$ and
$P_\bot=0$, one has to solve the Schr\"odinger equation:
\be\label{2}
P^-|m^2,\ P^+,\ P_\bot=0>=\frac{m^2}{2P^+}|m^2,\ P^+,\ P_\bot=0>  .
\ee
This approach appears promising in non-perturbative studies of gauge theories,
in particular QCD~\cite{cas,fra,per,ann}.

The quantization surface $x^+=0$, however, is a characteristic surface of the
field
equations. This peculiarity is reflected in infrared singularities, $P^+\to 0$,
in
such formulations. Consequently, one is forced to use some regularization.
Usually
the most simple regularization is chosen: $P^+\geq
\varepsilon >0$, where $\varepsilon$ is a cutoff parameter. The
simplicity of the vacuum and of the physical Fock space is related to this
choice
of the regularization.

The question about the equivalence of such a light-front formulation to the
usual
one arises. To answer this question,
results for various two-dimensional models have been considered: Sine
Gordon~\cite{ank,bur}, $\varphi^4$ model~\cite{har,wer}, QED~\cite{byl,ell},
QCD~\cite{hor}, etc. The results for the mass spectra agree rather well with
the
results of the usual approaches, except for some `vacuum effects'. These are
usually connected with condensates which are zero in the light-front formalism.
In
four-dimensional space-time the spectrum of positronium in QED was considered
with similar results~\cite{tan}.

To gain understanding about the equivalence of light-front formulation to the
ordinary one, it is useful to consider the theory again on a space-like plane,
close to the light front~\cite{pro,len}, and investigate the limiting
transition
to the latter. This can be done by introducing the following
coordinates~\cite{pro}:
\bey\label{3}
y^0&=&x^++\frac{1}{2}\eta^2 x^-,\nonumber\\
y^3&=&x^-,\nonumber\\
y^\bot&=& x^\bot,
\eey
with the metric $g_{\mu\nu}(\eta)\
(g_{0\nu}=0,g_{03}=g_{30}=1,g_{33}=-\eta^2)$.
The quantization plane is defined by $y^0=0$. The parameter $\eta$ is small and
in the
limit $\eta\to 0$ the exact  light front is approached.

In the studies~\cite{pro,len} of two-dimensional gauge theories formulated on a
finite $y^3$ interval with periodic boundary conditions, it was explicitly
shown that
one obtains equivalent results only, when the continuum
limit $L\to\infty$ is made first and then followed by the transition to the
exact
light cone
$\eta\to 0$ (or $L\eta\to\infty,\ \eta\to 0$). Taking the limit $\eta\to 0$ at
fixed
$L\ (L\eta\to 0)$ yields the usual light-front formulation (with $|x^-|\leq L)$
with
zero condensates.

Attempts have been made to take into account vacuum effects by considering zero
($P^+=0$)
Fourier modes of the fields~\cite{wer,pro,prk}. However, in
the light-front formulation these zero modes have peculiar
dynamics~\cite{fra,wer,prk}. For example, they depend on nonzero modes through
some
specific canonical constraints related to the choice of the boundary conditions
for $|x^-|\leq L$~\cite{fra,wer}. This means
that the physics at low momenta (zero modes) can depend on
high momentum modes in a complicated fashion.

In this paper another, more efficient approach to light-front quantization is
proposed. It is based on approximations for the vacuum in the ordinary
formulation \cite{stev}
and on the appropriate choice of canonical variables reflecting the
non-triviality of the vacuum in the given approximation. In terms of these new
variables we then take the naive light front limit $(\eta\to 0$ at fixed
$\varepsilon$); the resulting theory will include information on the
approximate,
non-trivial, vacuum.

This approach is demonstrated by two simple examples: scalar field
theory in two dimensions
(next section) and the massive Schwinger model (section 3).

\section{Scalar field theory in 1+1 dimensions}

For scalar field theory, we define the Lagrangian density as
\be\label{4}
{\cal
L}(\varphi)=\frac{1}{2}g^{\mu\nu}\partial_\mu\varphi(y)\partial_\nu\varphi(y)-
\frac{1}{2}m^2_0 \varphi^2(y)-\lambda U(\varphi),
\ee
where $U(\varphi)$ is an interaction term.  The theory is formulated using the
$y^\mu$
coordinates, eq.~(\ref{3}); here, in the two-dimensional case, the space
coordinate
is denoted by $y^1$. Consequently, we can write
\be\label{5}
{\cal
L}(\varphi)=\partial_0\varphi(y)\partial_1\varphi(y)+\frac{1}{2}\eta^2(\partial_0
\varphi(y))^2-\frac{1}{2}m^2_0\varphi^2(y)-\lambda U(\varphi).
\ee
After introducing the canonical variable $\Pi(y)$, the conjugate momentum of
$\varphi(y)$,
\begin{displaymath}
\Pi(y)=\frac{\partial
L}{\partial(\partial_0\varphi(y))}=\eta^2\partial_0\varphi(y)
+\partial_1\varphi(y),
\end{displaymath}
the Hamiltonian reads
\bey\label{6}
H&=&\int dy^1\left\{\frac{(\Pi-\partial_1\varphi)^2}{2\eta^2}+\frac{1}{2} m^2_0
\varphi^2+\lambda U(\varphi)\right\}\nonumber\\
&=&:\ H_0+\lambda U.
\eey
The usual (equal $y^0$) commutation relations  are imposed:
\begin{displaymath}
[\varphi(y^1),\Pi(y^{1\prime})]=i\delta(y^1-y^{1\prime}).
\end{displaymath}
We make a Fourier-decomposition of the canonical fields $\varphi$ and $\Pi$ in
terms of
the ``bare'' operators $b$ and $b^+$ ($b|0_b>=0$, with $|0_b>$
as the free field vacuum):
\bey\label{7}
\varphi(y)&=&\frac{1}{\sqrt{4\pi}}\int^{+\infty}_{-\infty}
\frac{dk_1}{\sqrt{E_0(k_1)}}(b(k_1)+b^+(-k_1))e^{-ik_1y^1}\nonumber\\
\Pi(y)&=&\frac{-i}{\sqrt{4\pi}}\int^{+\infty}_{-\infty}dk_1\sqrt{E_0(k_1)}
(b(k_1)
-b^+(-k_1))e^{-ik_1y^1},
\eey
where $E_0(k_1)=\sqrt{k^2_1+\eta^2 m^2_0}$. For the operators $b$ and $b^+$ we
have standard commutation relations
\begin{eqnarray*}
&&[b(k_1),b^+(k'_1)]=\delta(k_1-k'_1)\\
&&[b(k_1),b(k'_1)]=0 = [b^+(k_1),b^+(k'_1)].
\end{eqnarray*}
In terms of $b$ and $b^+$ $H_0$ is diagonal by construction:
\be\label{8}
H_0=\int^{+\infty}_{-\infty} dk_1\frac{E_0(k_1)-k_1}{\eta^2}b^+(k_1)b(k_1).
\ee
Since $\eta$ appears also in the energy $E_0$, only terms in $H_0$ with
$k_1\leq0$
are singular in the limit $\eta\to 0$. In order to make the energy finite in
this
limit we consider the restricted Fock space ${\cal F}_{(\varepsilon)}$
\begin{displaymath}
{\cal F}_{(\varepsilon)}=:\ \{\prod_i b^+(k_i)|0_b>, \ k_i\geq\varepsilon>0\},
\end{displaymath}
where $\varepsilon$ is the cutoff parameter. If we now take $\eta\to 0$ at
fixed
$\varepsilon>0$, we obtain a finite result for the energy, because
$\frac{E_0(k_1)-k_1)}{\eta^2}\to\frac{m^2_0}{2k_1}$, for $k_1\geq\varepsilon>0$
and
$\eta\to 0$.
The limiting form of the Hamiltonian on the subspace ${\cal F}_{(\varepsilon)}$
reproduces the usual light-cone   Hamiltonian $P^-$
\bey\label{9}
P^-&=&\lim_{\eta\to 0}\ H_\eta\qquad\mbox{(acting on } {\cal
F}_{(\varepsilon)})
\nonumber\\
&=&\int^{+\infty}_{-\infty} dx^-\left\{\frac{1}{2} m^2_0 \varphi^2_\varepsilon
(x)+\lambda U(\varphi_\varepsilon)\right\},
\eey
where $\varphi_\varepsilon(x)$ is the parametrization of the field in
light-front
coordinates
\be\label{10}
\varphi_\varepsilon(x^-,x^+=0)=\frac 1 {\sqrt{4\pi}}
\int^\infty_\varepsilon\frac
 {dp^+}
{\sqrt{p^+}}\left[ b(p^+)e^{-ip^+x^-}+b^+(p^+)e^{ip^+x^-}\right]  .
\ee
Note that we would get the same result in the theory formulated on a finite
interval,
$-L\leq y^1\leq L$, with periodic boundary conditions in $y^1$. In this case
the role
of the cut-off parameter $\varepsilon$ would be taken over by the parameter
$\pi/L$.
Furthermore, the result (\ref{9}) can be obtained via time-independent
perturbation
theory in $\eta$~\cite{pro},~\cite{len}.

At this point we want to introduce a better way to formulate the light-cone
limit
$\eta\to 0$. Before the limiting transition, i.e. still for finite $\eta$, we
approximate the vacuum by a Gaussian trial state~\cite{stev,pol} using the
limit $\varepsilon\to0$.
This trial state is parametrized by a Bogoljubow type transformation:
\be\label{11}
|0_a>=\exp\left[-\frac{1}{2}\int dk f(k) (b^+(k)
b^+(-k)-b(k)b(-k))+f_0(b^+(0)-b(0))
\right] |0_b> ,
\ee
where $f(k)$ and $f_0$ are real, and $f(k)=f(-k)$.  The trial vacuum can be
easily
defined with new operators $a(k_1),a^+(k_1)$ such that $a(k_1)|0_a>=0$.
As follows from eq. (\ref{11})
these new operators $a(k_1)$ and $a^+(k_1)$ are linear combinations of the old
operators  $b^+(k_1)$ and $b(k_1)$. Therefore one can rewrite the
Fourier-decompositions of $\varphi$ and $\Pi$ in terms of $a,a^+$:
\bey\label{12}
\varphi(y)&=&\varphi_0+\frac{1}{\sqrt{4\pi}}\int\frac{dk_1}{\sqrt{E(k_1)}}
(a(k_1)+a^+(-k_1))e^{-ik_1y^1}    ,
\nonumber\\
\Pi(y)&=&\frac{-i}{\sqrt{4\pi}}\int
dk_1\sqrt{E(k_1)}(a(k_1)-a^+(-k_1))e^{-ik_1y^1}   .
\eey
Identifying these expressions with the corresponding ones
in terms of the $b$ and $b^+$
operators (eq. (\ref{7}))  yields the linear transformations between the sets
$(a,a^+)$
 and  $(b,b^+)$ in terms of $E(k_1), E_0(k_1)$ and $\varphi_0$. Then the
condition
$a(k_1)|0_a>=0$ determines the relation between $(E(k_1),\varphi_0)$ and
$(f(k),f_0)$
to be:
\bey\label{13}
E(k_1)&=& E_0(k_1)\exp(2f(k))\nonumber\\
\varphi_0&=&\frac{f_0}{\sqrt{\pi\eta m_0}}\frac{1-\exp(-f(0))}{f(0)}.
\eey
In the following we will consider $E(k_1)$ and $\varphi_0$ as parameters  of
the
transformation (or, equivalently, of the trial state).

 From now on we specify the interaction as $U(\varphi)=\varphi^4$. We proceed
by
rewriting  the Hamiltonian $H$ in the normal ordered form with respect to the
$a,a^+$
operators; :: denotes this normal ordering. The result is:
\bey\label{14}
H&=&:\int dy^1\left[\frac{(\Pi-\partial_1\varphi)^2}{2\eta^2}+\frac{1}{2}
(m^2_0+12\lambda
\underbrace{\varphi\varphi})\varphi^2+\lambda\varphi^4+\frac{1}{2\eta^2}
\underbrace{\Pi\Pi}\right.\nonumber\\
&&\left.+\frac{1}{2\eta^2}\underbrace{\partial_1\varphi\ \partial_1\varphi}
+\frac{1}{2} m^2_0\underbrace{\varphi\varphi}
+3\lambda(\underbrace{\varphi\varphi})^2\right] :,
\eey
where
\begin{eqnarray*}
&&\underbrace{\varphi\varphi}:=\int\frac{dk_1}{4\pi E(k_1)}\qquad\qquad
\underbrace{\Pi\Pi}:=\int\frac{dk_1}{4\pi} E(k_1)\\
&&\underbrace{\partial_1\varphi\ \partial_1\varphi}:=\int\frac{dk_1(k_1)^2}
{4\pi E(k_1)}.
\end{eqnarray*}
These integrals are understood to be regularized by a cut-off parameter
$\Lambda$,
$|k_1|<\Lambda$. In order to fix the parameters $E(k_1)$ and $\varphi_0$ we
minimize
the expectation value of the Hamiltonian density ${\cal H}$ in the trial vacuum
$|0_a>$.
This expectation value is given by
\begin{eqnarray*}
<0_a|{\cal H}|0_a>&=&\frac{1}{2\eta^2}(\underbrace{\Pi\Pi}
+\underbrace{\partial_1\varphi\
\partial_1\varphi})+\frac{1}{2}(m^2_0+12\lambda\varphi^2_0)
\underbrace{\varphi\varphi}\\
&&+3\lambda(\underbrace{\varphi\varphi})^2+\frac{1}{2}
m^2_0\varphi^2_0+\lambda\varphi^4_0\\
&=&\frac{1}{8\pi\eta^2}\int dk_1(E(k_1)+\frac{k^2_1+\eta^2
m^2_0+12\eta^2\lambda\varphi^2_0}{E(k_1)})\\
&&+3\lambda\left(\int\frac{dk_1}{4\pi E(k_1)}\right)^2+\frac{1}{2}
m^2_0\varphi^2_0
+\lambda\varphi^4_0.
\end{eqnarray*}
At the extremum, $\frac{\delta<0_a|{\cal H}|0_a>}{\delta E(k_1)}=0\ , \frac{
\delta<0_a|{\cal H}|0_a>}{\delta\varphi_0}=0$, we obtain:
\bey\label{16}
E^2(k_1)&=& k^2_1+\eta^2\left[m^2_0+12\lambda\varphi^2_0+\frac{3\lambda}{\pi}
\int_{|q_1|\leq \Lambda} dq_1 E^{-1}(q_1)\right]\nonumber\\
:&=& \  k^2_1 +\eta^2 m^2,
\eey
and
\be\label{17}
\varphi_0(m^2-8\lambda\varphi^2_0)=0.
\ee
Using the equality (\ref{16}) we get for a large cut-off $\Lambda$:
\[  \int_{|q_1|\leq\Lambda}dq_1 E^{-1}(q_1)
\begin{array}{c} {} \\ \approx \\ (\Lambda/m)\to\infty
\end{array}\ln\frac{4\Lambda^2}{\eta^2
m^2}=\ln\frac{4\Lambda^2}{\eta^2\lambda}+\ln\frac{\lambda}{m^2}\approx\ln
\frac{4\Lambda^2}{\eta^2\lambda}.
\]
Since  in this limit $m^2\simeq m^2_0+12\lambda\varphi_0^2+\frac{3\lambda}{\pi}
\ln\frac{4\Lambda^2}{\eta^2\lambda}$, we can renormalize the theory by
choosing:
\[ \frac{m^2_0}{\lambda}=\frac{3}{\pi}\ln\frac{\eta^2\lambda}{4\Lambda^2}+\xi,
\]
with  a parameter $\xi, \ -\infty<\xi<\infty$. Then we can convert equation
 (\ref{16})
into a nonlinear equation for $m$:
\bey\label{18}
&&y+\frac{3}{\pi}\ln y=\xi+12\varphi^2_0,\nonumber\\
&&\mbox{with}\ y=m^2/\lambda.
\eey
This equation should be solved together with eq.~(\ref{17}), which obviously
has the
solutions:
\bey\label{19}
&&(1)\quad\varphi_0=0\nonumber\\
&&(2)\quad\varphi_0^2=m^2/8\lambda=\frac{1}{8} y.
\eey
Therefore, there are two different cases:
\bey
&&(1)\qquad y+\frac{3}{\pi}\ln y=\xi\nonumber\\
\mbox{and}\qquad && \label{20}\\
&&(2)\qquad -\frac{1}{2} y+\frac{3}{\pi}\ln y=\xi.\nonumber
\eey
The solutions $y_1(\xi)$ (full curve), $y_2(\xi)$ (dashed curve)
are shown in Fig. 1. Of
course, one needs to choose the solution which corresponds to a minimum of the
(trial)
vacuum energy. The difference of this energy in the cases (1) and (2)
can be calculated
straightforwardly; the result is
\[ <0_a|{\cal H} | 0_a>_{(1)}-<0_a|{\cal
H}|0_a>_{(2)}=\frac{\lambda}{8\pi}(y_1-y_2)+\frac{\lambda}{48}\left(
y^2_1+\frac{1}{2} y^2_2\right).
\]
At the
critical point $\xi_c=-0.503...$,  the sign of the energy difference changes
and,
consequently, the favoured solution switches from $y_2$ to $y_1$ (for
increasing $x$).
In other words, we obtain the well-known phase transition in this approximation
\cite{stev,pol}. Moreover, the exact location of the phase transition, i.e. the
critical
point, agrees with earlier results \cite{stev} .

For the minimal energy solution of eqs. (\ref{18}) and (\ref{19}) we introduce
the
notation
\[ F(\xi)\ :=\left\{ \begin{array}{ll}
\left(\frac{m^2}{\lambda}\right)_1&,\quad\xi>\xi_c\\ \\
\left(\frac{m^2}{\lambda}\right)^{up}_2&,\quad\xi<\xi_c ,\end{array}\right. \]
where $\left(\frac{m^2}{\lambda}\right)^{up}_2$ denotes the upper branch of the
curve
$\left(\frac{m^2}{\lambda}\right)_2$. Now we are in the position to present a
renormalized Hamiltonian, which is obtained by subtracting the trial-vacuum
energy,
\bey\label{20a}
H_{ren}&=&:\ \int
dy^1\left[\frac{(\Pi-\partial_1\tilde\varphi)^2}{2\eta^2}+\frac{1}{2}\lambda
F(\xi)\tilde\varphi^2+\right.\nonumber\\
&&\left.\lambda \sqrt{2F(\xi)} \theta
(\xi_c-\xi)\tilde\varphi^3+\lambda\tilde\varphi^4 \right]\ :\ ,
\eey
with $\tilde\varphi\  := \varphi-\varphi_0$.

We can use $H_{ren}$ as the starting Hamiltonian for the
limiting transition to the light-cone. Repeating the steps following eq.
(\ref{8}), we
obtain the effective light-front Hamiltonian
\be\label{21}
P^-\ =\ :\int dx^-\left[\frac{\lambda}{2} F(\xi)\tilde\varphi^2_\varepsilon(x)
+\lambda\sqrt{2F(\xi)}\theta(\xi_c-\xi)\tilde\varphi^3_\varepsilon(x)
+\lambda\tilde\varphi^4_\varepsilon(x)\right]:.
\ee
This expression differs from the usual one~\cite{har} by the presence of the
function
$F$  describing vacuum effects. In the quadratic term,
we see that the effective theory has a renormalized mass term.
The cubic term was even completely absent in the
usual approach. For $\xi>\xi_c$, i.e. in the phase without zero mode
($\varphi_0=0$),
the cubic term
vanishes identically and the mass renormalization is all that remains. For
$\xi<\xi_c$, the reflection symmetry $\varphi\to-\varphi$ is spontaneously
broken
and a zero mode $\varphi_0\not=0$ is present. This zero mode produces in the
effective light-cone Hamiltonian an additional interaction term, which
explicitly breaks the reflection symmetry. In other words, this
formulation converts a spontaneous symmetry breaking into an explicit
symmetry breaking in the effective light-cone Hamiltonian. In this way, a
rather
long-standing defect of light-cone quantization, namely the triviality of the
vacuum,
can be handled in an approximative way. We emphasize that the proposed
 approach is very
reasonable. The zero modes carry infinite light-cone energy. The strategy to
remove
high-energy degrees of freedom by effective interactions is the usual strategy
of
renormalization in equal time field theory.

The effective light-cone Hamiltonian, eq. (\ref{21}),  can  be used for
explicit calculation using standard light-cone techniques. We
 note that this approach can easily be generalized to other scalar field
theories
in two or more dimensions.

\section{Massive Schwinger Model}

The massive Schwinger has also been formulated in the $y^\mu$ coordinates, i.e.
for
$\eta\not= 0$ \cite{pro,len}. The Lagrangian density reads
\bey\label{22}
{\cal L}(A_\mu,\psi)&=&-\frac{1}{4} g^{\mu\rho} g^{\nu\lambda} F_{\mu\nu}(y)
F_{\rho\lambda}(y) \nonumber\\
&&+\bar\psi(x(y))[i\left(\frac{\partial y^\lambda}{\partial
x^\mu}\right)\gamma^\mu
D_\lambda -M]\psi(x(y)),
\eey
where the covariant derivative,
\begin{displaymath}
D_\mu=\partial_\mu-ieA_\mu(y),
\end{displaymath}
and the field strength tensor,
$F_{\mu\nu}=\partial_\mu A_\nu-\partial_\nu A_\mu$, are expressed in terms of
the
 vector
potential $A_\mu$. The fermion field contains two spinor components,
$\psi=\left(
\begin{array}{c} \psi_+\\ \psi_-\end{array}\right)$, and $M$ is the fermion
mass.
With the definition (\ref{3}) of the coordinates and the $\gamma$-matrices,
\begin{displaymath}
\gamma^0=\left(\begin{array}{cc} 0&-i\\i&0\end{array}\right) , \gamma^1=\left
(\begin{array}
{cc} 0&i\\ i&0\end{array}\right),
\end{displaymath}
we obtain for the Lagrangian density
\bey\label{23}
{\cal L}(y)&=&\frac{1}{2} F^2_{01}(y)+i\sqrt2 \psi^\dagger_+
D_0\psi_++\frac{1}{2}
\sqrt2 i\eta^2
\psi^\dagger_-  D_0\psi_-\nonumber\\
&&+i\sqrt2\psi^\dagger_-D_1\psi_-
-iM(\psi^\dagger_-\psi_+-\psi_+^\dagger\psi_-).
\eey
Note that only the mass term couples the two fermion components $\psi_-$
and $\psi_+$. Let
us consider the theory on a finite $y^1$ interval: $-L\leq y^1\leq L$ and
impose periodic
boundary conditions on the fields $A_\mu$ and $\psi$. We fix the gauge by
imposing
\be\label{24}
\partial_1 A_1=0,
\ee
i.e. the `Coulomb gauge'.

It has been shown earlier \cite{pro,len} that the zero mode of $A_1$ cannot be
gauged
away.  In the Coulomb gauge the only constraint is Gauss's law
\begin{displaymath}
\partial_1 F_{01}+e\sqrt2 \psi^\dagger_+\psi_++\frac{1}{2} \sqrt 2
 e\eta^2\psi_-^\dagger\psi_-=0.
\end{displaymath}
It can be solved with respect to the nonzero modes of $F_{01}$:
\be\label{25}
\left\langle F_{01}\right\rangle =-\partial^{-1}_1(e\sqrt2 \psi^\dagger_+
\psi_+
+\frac{1}{2}\sqrt2 e\eta^2 \psi^\dagger_-\psi_-),
\ee
where the special brackets $\bigl\langle\ \bigr\rangle$ define the non-zero
modes
$\bigl\langle f\bigr\rangle =f(y^1)-\frac{1}{2L}\int^L_{-L} f(y^1)dy^1$
and $\partial^{-1}_1$ is the periodic Greens function of the differential
operator
$\partial_1$ (see e.g. \cite{Len1}),
\be\label{25a}
\partial^{-1}_1 (x)= \sum_{n \neq 0} \frac{1}{2i\pi n} \exp (2i\pi n
\frac{x}{L}).
\ee
 Substituting eq. (\ref{25}) into the Lagrangian
 and performing the Legendre transformation yields the Hamiltonian
 in terms of
the canonical variables $\chi_+=\eta^{1/4}\psi_+,\
 \chi_-=\eta^{-1/4}\eta\psi_-,\
\Pi_1=\int^L_{-L} dy^1 F_{01}(y^1)$ and $A_1$
\bey\label{27}
H&=&\int^L_{-L}dy^1\left\lbrace\frac{\Pi^2_1}{8L^2}+\frac{1}{2} e^2\left
(\partial^{-1}_1
(\chi^\dagger_+\chi_++\chi^\dagger_-\chi_-)\right)^2\right.
\nonumber\\
&&\left.
-2i/\eta^2\chi^\dagger_-
D_1\chi_-+i\frac{M}{\eta}(\chi^\dagger_-\chi_+-\chi^\dagger_+
\chi_-)
\right\rbrace.
\eey
Moreover, integration of Gauss's law gives a residual constraint, which is to
be
imposed on the physical  states
\be\label{27a}
\int^L_{-L} dy^1 (\chi^\dagger_+\chi_++\chi_-^\dagger\chi_-)|phys>=0.
\ee
Notice that our canonical variables satisfy the following commutation
relations:
\begin{eqnarray}\label{29}
&&\left\{\chi^\dagger_\pm
(y^1),\chi_\pm(y^{1'})\right\}_{y^0=y^{0'}}=\delta(y^1-y^{1'}),\nonumber\\
&&[A_1(y^0),\pi_1(y^0)]=i.
\end{eqnarray}
The regularized  charge densities of the right and left movers are obtained via
point splitting the two densities and connecting the two centers with a string.
\be\label{30}
I_\pm (y^1)=\lim_{\varepsilon\to 0}\left(\chi^\dagger_\pm \left( y^1\mp
\frac{i\varepsilon}{2}\right)\chi_\pm \left(y^1\pm\frac{i\varepsilon}{2}
\right)\exp(\pm\varepsilon eA_1)-\frac{1}{2\pi\varepsilon}\right)
\ee
$$
:=\lim_{\varepsilon\to 0}\left(I_\pm (y^1,\varepsilon)-
\frac{1}{2\pi\varepsilon}\right)  .
$$
These chiral charge densities have Fourier expansions $(r=\pm,\ p_n=\pi n/L)$:
\be\label{31}
I_r(y)=\frac{1}{2L}\left( Q_r+\sum_{n\not= 0}\sqrt{|n|} I_{n,r}
(y^0)\exp(-ip_n y^1)\right)   .
\ee
The zero mode part of the Fourier expansion is defined by the total chiral
charges
\be\label{32}
Q_r=\int^{+L}_{-L} I_r(y^1)dy^1   ,
\ee
which can be calculated via the $\varepsilon$ prescription by inserting
the $\varepsilon$-regularized charge density $I_r(y^1,\varepsilon)$
into eq.  (\ref{32})
\begin{eqnarray}
Q_r=\lim_{\varepsilon\to 0}(Q_r(\varepsilon)&-& L/\pi\varepsilon).
\label{32a}
\end{eqnarray}
The coefficients $I_{n,r}$ obey commutation relations, which are a consequence
of the
commutation relations of the Fourier coefficients
$\chi_{n,r}$ of the fermion fields and of the regularization, eq. (\ref{30}):
\bey
\chi_r(y^1)&=&\frac{1}{\sqrt{2L}}\sum^\infty_{n=-\infty}\chi_{n,r}(y^0)
\exp(-ip_ny^1);\\ \label{33}
\left\{ \chi_n(y_0),\chi^\dagger_{n'}(y_0)\right\} &=&\delta_{nn'}\\ \label{34}
[I_{n,r},I^\dagger_{n'r'}] &=& r\frac{n}{|n|}\delta_{rr'}
\delta_{nn'};\ n\not=0.\label{35}
\eey
As usually \cite{man,prh}, we define subspaces $|\ell>$ of the total Hilbert
space, which correspond
to sectors ($\ell=(\ell_+,\ell_-)$) with given edges of occupied energy levels
for the right and left movers
as follows:
\bey\label{36}
\chi_{n,r}|\ell>&=&\theta(r\ell_r-rn)|\ell> , \nonumber\\
\mbox{with}\ \theta(\ell)=\left\{\begin{array}{cc}
1&\ell>0\\
0&\ell\leq 0\  \end{array}\right.
\eey
Consequently, the operators $I_{n,+},I^\dagger_{n,-},n>0$, annihilate the
states $|\ell>$:
\be\label{36'}
I_{n,+}|\ell>=I^\dagger_{n,-}|\ell>=0  .
\ee
The charge eigenvalues in sectors $|\ell>$ depend on the zero mode gauge field
\cite{man,prh}
\be\label{37}
Q_r|\ell>=(r\ell_r+r\frac{eLA_1}{\pi} +1/2)|\ell>,\ \ \
[Q_{r},\Pi_1]=\frac {rieL} \pi  .
\ee
We introduce~\cite{prh} the variables $\omega_r$ canonically conjugated to the
$Q_r$ such that
\be\label{38}
[\omega_r,Q_{r'}]=i\delta_{rr'}   .
\ee
Then we can represent the fermion fields with the help of the bosonic operators
$I_r,\omega_r,A_1$\cite{prh,byll,iso}:
\pagebreak
\bey\label{39}
\chi_r(y^1)=\frac{1}{\sqrt{2L}}&\exp&(-i\omega_r)\cdot\nonumber\\
&\exp&\left(\frac{ir\pi}{2}(Q_++Q_--1)-\frac{ri\pi y^1}{L}\left(Q_r
-\frac{reLA_1}{\pi}-\frac{1}{2}\right)\right)\cdot\nonumber\\
&\exp&\left(-\sum_{n>0}\sqrt{n}I^+_{n,r}e^{irp_ny^1}\right)\cdot
\exp\left(+\sum_{n>0}\sqrt{n} I_{n,r}e^{-irp_ny^1}\right).
\eey

These operators satisfy the commutation relations eq. (\ref{29}) and reproduce
the
regularized charge densities, eq. (\ref{29}). The necessary explanations can be
found in the Appendix.
The operators $I_r$ link the fermionic  to the bosonic description:
In the Hamiltonian of eq. (\ref{27}) we recognize  four terms. The first three
terms
can be rewritten as a free boson Hamiltonian in terms of bosonic variables
$(\phi,\Pi_\phi)$ constructed from
the charge densities $I_r$.
\bey\label{40}
\Pi_\phi&=&\sqrt\pi (I_+-I_-) , \nonumber\\
\phi&=&-\frac{1}{m}\left(\frac{\Pi_1}{2L}-\partial^{-1}(
e(I_++I_-))\right) , \nonumber\\
m^2&=&e^2/\pi.
\eey
With the help of the commutation relations (\ref{34}) one can verify that
$\Pi_\phi$
and $\phi$ are canonically conjugate variables. The mass term of the bosonic
Hamiltonian is easily calculable using the fact that the zero mode is
subtracted in
$\langle e(I_++I_-)\rangle$
\be\label{41}
\int^L_{-L}dy^1\frac{1}{2} m^2\phi^2=\int^{+L}_{-L} dy^1\left[\frac{1}{8L^2}
\Pi^2_1+\frac{e^2}{2}(\partial_1^{-1}( I_++I_-))^2\right]  .
\ee
The momentum term can be expressed with the help of eq. (\ref{39})  in terms of
the
chiral charges. The space integral of the square of the zero mode free chiral
charge
density $\langle I_r\rangle^2$ is related to the fermionic momentum
\bey
\pi\int^{+L}_{-L}dy^1\langle &I_r&\rangle ^2=r\int^{+L}_{-L}dy^1\chi^\dagger_r(
(y^1)(iD_1)\chi_r(y^1)  .\label{42}
\eey
This relation is derived in the Appendix. On the physical subspace defined by
\be\label{43a}
Q|phys>=0, \,\, Q=Q_++Q_-   ,
\ee
we obtain with eqs. (\ref{40}, \ref{42}):
\be\label{43}
\int^{+L}_{-L} dy^1\frac{(\Pi_\phi-\partial_1\phi)^2}{2\eta^2}=\frac{2}{\eta^2}
\int^{+L}_{-L}dy^1\chi^\dagger_-(y^1)(-iD_1)\chi_-(y_1).
\ee
The mass term remains as last term in the fermionic Hamiltonian of
eq. (\ref{27}). It is given by  direct insertion of the boson representation
of the fermion fields  eq.  (\ref{39}) into eq. (\ref{27}).
After simplifying this expression with the help of normal ordering
with respect to $I^{\dagger}_{n,r}$ and $I_{n,r}$ (cf. Appendix) we obtain:
\be\label{44}
\frac{iM}{\eta}\int^{+L}_{-L}dy^1\left[\chi^\dagger_-(y^1)\chi_+(y^1)-\chi^\dagger_+(y^1)
\chi_-(y^1)\right]
\ee
$$
=-\frac M {\eta L} :\sin(\omega_+ -\omega_- +
\sqrt{4\pi}\langle\phi\rangle): .
$$

In a similar way to the treatment of the  scalar field theory in $(1+1)$
dimensions we approximate the vacuum by a trial state $|0_a>$ which is defined
as
\be\label{45}
a_n|0_a>=0,
\ee
where $a_n$ and $a_n^+$ are the normal modes of the boson variables
$\phi(y^1),\Pi_\phi(y^1)$
\bey\label{46}
\phi(y^1)&=&\frac{1}{\sqrt{2L}}\sum^\infty_{n=-\infty}\frac{1}{\sqrt{2E_n}}
(a_n+a_{-n}^+) e^{-i p_n y^1},\nonumber\\
\Pi_\phi(y^1)&=&\frac{-i}{\sqrt{2L}}\sum^{\infty}_{n=-\infty}\sqrt{\frac{
E_n}{2}}(a_n-a_{-n}^+)
e^{-i p_n y^1}.
\eey
The weights $E_n$ are variational parameters, which then also enter the Fourier
coefficients of the chiral charges (cf. eq. (\ref{40}))
\be\label{47}
I_{n,r}=\frac{-ri}{\sqrt{4E_n|p_n|}}\left[(E_n+rp_n)a_n-(E_n-rp_n)
a^+_{-n}\right].
\ee
Inserting these expressions into eq. (\ref{44}) and normal ordering with
respect to
the trial vacuum  eq. (\ref{45}) we obtain on the physical subspace,
eq. (\ref{43a}), the effective Hamiltonian
\bey\label{48}
{\cal H}&=&\int^L_{-L}dy^1\left[\frac{(\Pi_\phi-\partial_1\phi)^2}{2\eta^2}+
\frac{1}{2}m^2\phi^2\right.\nonumber\\
&&\left.-\frac{M}{\eta L}\exp\left\{\frac{\pi}{L}\sum_{n>0}\left(\frac{1}{p_n}
-\frac{1}{E_n}\right)\right\}\ :\cos(\omega+\sqrt{4\pi}\langle\phi\rangle
):\right],
\eey
with $\omega=\omega_+-\omega_--\frac {\pi} 2$. Note that the normal
ordering symbol : : means here to order the operators $a_n,a^\dagger_n$.

In order to fix the variational parameters we look for a minimum of the vacuum
energy density using the trial vacuum state $|0_a>$. The calculation proceeds
in an
analogous way to the calculation of the $\varphi^4$ scalar theory. Note
that the minimum corresponds to $\omega=0$. Using condition (\ref{46}) we
obtain the
following expression for the vacuum energy density of the Hamiltonian, eq.
(\ref{49}), at $\omega=0$:
\bey\label{49}
<0_a|{\cal H}(y)|0_a>&=&\frac{1}{4L\eta^2}\sum_{n>0}\left(
E_n+\frac{p^2_n+\eta^2
m^2}{E_n}\right)\nonumber\\
&&-\frac{M}{\eta L}\exp\left(\frac{\pi}{L}\sum_{n>0}\left(\frac{1}{p_n}
-\frac{1}{E_n}\right)\right).
\eey
At the minimum of this expression we have $(n>0)$:
\be\label{50}
E^2_n=p^2_n+\eta^2 m^2+\frac{4\pi M\eta}{L}\exp\left\{\frac{\pi}{L}
\sum_{n>0}\left(\frac{1}{p_n}-\frac{1}{E_n}\right)\right\}=p^2_n+\eta^2\mu^2,
\ee
with \ $\mu^2=m^2+\frac{4\pi M}{\eta L}\exp\left\{\frac{\pi}{L}\sum_{n>0}
\left(\frac{1}{p_n}-\frac{1}{E_n}\right)\right\}$.

 From this equation $\mu$ is to be determined. In order to do that we rewrite
the
infinite sum in the exponent \cite{grad} as
\[ \frac{\pi}{L}\sum_{n>0}\left(\frac{1}{p_n}-\frac{1}{E_n}\right)
=-2\sum^\infty_{k=1}K_0(2\pi ak)+\gamma+\ln \frac{1}{2} a+\frac{1}{2a},
\]
where we introduced $a=\frac{L\eta\mu}{\pi}$; $K_0$ is the modified
Besselfunction
and $\gamma=0.5772...$ (Euler's constant). In the limit $\eta m L\gg 1,\
a\gg1$, the sum gives $\gamma+\ln\frac{1}{2} a$ and one readily obtains
\be\label{51}
\mu^2=m^2+\frac{4\pi M}{\eta L}(\frac{1}{2} a e^\gamma)=m^2+2e^\gamma M\mu,
\ee
which gives
\bey\label{52}
\mu&=&e^\gamma(M+\sqrt{M+e^{-2\gamma}m^2})\nonumber\\
&=&(M_\gamma+\sqrt{M_{\gamma}^2+ \frac{e^2}{\pi}}),
\; \; \; M_\gamma \equiv e^\gamma M
\eey
This value of $\mu$ corresponds to the effective boson mass parameter in the
Hamiltonian  (\ref{49}). Some remarks are in order. Taking the $L\to\infty$
corresponds to $\varepsilon\to 0 ,(|k_1|\geq\varepsilon)$  in the scalar
field theory  (eqs. (\ref{16})  and (\ref{17})). For obtaining the effects of
the
non-trivial vacuum one needs to take these limits in the relevant equations
(cf.
eqs. (\ref{16}), (\ref{17}) and (\ref{18})). Indeed, immediately approaching
the
light  front, $\eta\to 0$ at finite $L$ would not  reproduce the boson mass,
eq.
(\ref{52}).  This can easily be seen from eq. (\ref{51}) in combination with
the
small $a$ limit of the infinite sum:
$\frac{\pi}{L}\sum_{n>0}\left(\frac{1}{p_n}
-\frac{1}{E_n}\right) =
\frac{1}{2} a^2\xi(3) + O(a^3)$, where $\xi$ is the Riemann function;
$\xi(3)=1.201056903...$.
Actually, in this limiting case $\mu$ diverges as $\frac{1}{\sqrt{\eta}}$.

The next step is to take this Hamiltonian, eq. (\ref{49}), with $E_n$ and $\mu$
fixed by eqs. (\ref{51}) and (\ref{52}), as the starting one for the transition
to
the light-cone formulation. Repeating the procedure as outlined in section 2
for
the scalar field theory, we obtain the following effective light-cone
Hamiltonian
for the massive Schwinger model

\be\label{53}
P^-=\ :\int^L_{-L}dx^-\left\{\frac{e^2}{\pi}\varphi_L^2+\frac{M_\gamma}{2\pi}
(M_\gamma+\sqrt{M_\gamma^2+
e^2/\pi})[1-\cos\sqrt{4\pi}\varphi_L]\right\}\ :,
\ee
where $\varphi_L(x)$ is the analog of light-cone field, defined by the Fourier
decomposition
\be\label{54}
\varphi_L(x)=\sum_{n>0}\frac{1}{\sqrt{4L p_n}}(a_n e^{-ip_n x^-}+a^+_n e^{i p_n
x^-}).
\ee
Note that the operators $a_n,a_n^+, n>0$ form the light-cone Fock basis.
The field $\varphi_L(x)$ can be expressed in terms of light
front fermionic variables:
\[  \varphi_L(x)=-\sqrt\pi \partial^{-1}_-(
\chi^+_+\chi_+).\]
This means that the expression (\ref{54}) can be written on the light cone
also in the fermionic basis.

It should be emphasized that the result, eq. (\ref{54}), indeed yields a
correction
to the naive light-cone approach \cite{cole}.  In the future we hope to address
the
interesting question how this affects the mass spectrum, in particular for
small
fermion mass.

The generalization of this approach to gauge theories in higher dimensions may
be
attempted with the help of Hamiltonians where the dependent degrees of freedom
have
been eliminated after gauge fixing \cite{Len1,Len2}.

\section {Acknowledgments}
This work was supported by the exchange program between the University of
Heidelberg and the University of St. Petersburg. We would like to thank Profs.
F.
Lenz and H.C. Pauli, as well as Th. Gasenzer for useful discussions.

\begin{appendix}
\renewcommand{\theequation}{\Alph{section}.\arabic{equation}}
\setcounter{equation}{0}

\section{Appendix}
In this appendix we give clarifications of eqs. (\ref{39}), (\ref{42}) and
(\ref{44}), based on the
more general considerations of \cite{prh,byll}.
Let us demonstrate that the expression in eq.
(\ref{39}) satisfies  canonical anticommutation relations.
First of all, notice  that the
representation (\ref{39}) acts in the Hilbert space spanned by
vectors of the form
$\prod_{i,j}\lbrace I^+_{n_i,+}\rbrace
\lbrace I_{n_j,\cdot}\rbrace|\ell\rangle,\  n_i>0,\ n_j>0$,
where the $I_{n,+},I^+_{n,-}$ and $I^+_{n,+},I_{n,-}$ act like
annihilation and creation operators
with  respect to ``vacuum" states $|\ell\rangle$ according to
eqs. (\ref{36}), (\ref{36'}) at $n>0$.
Using for these operators the normal ordering symbol ::
we can rewrite eq. (\ref{39}) in more
compact form:
\be
\chi_r(y^1)=\frac{1}{\sqrt{2L}}\exp(-i\omega_r)\exp\left\{ri\pi\left(\frac{1}{2}
\overline Q_r\right)\right\}:\exp\{-r2\pi i\partial^{-1}_1( I_r
)_{y^1}\}:,
\ee
where we denote by $\overline Q_r$ and $\overline Q$ the integer valued parts
of the
charges $Q_r$ and $Q$ ($\overline Q_r=Q_r-\frac{reLA_1}{\pi}-\frac{1}{2}$).

Let us consider the products $\chi_r(y^1)\chi^{\dagger}_{r'}(y^{1'})$
and $\chi^{\dagger}_{r'}
(y^{1'})\chi_r(y^1)$ as a function of $z=\exp(ri\pi y^1L^{-1}),z'=\exp(r'i\pi
y^{1'}L^{-1})$, taking the operator products in normal ordered form. We get
\bey
&&
\chi_r(y^1)\chi^{\dagger}_{r'}(y^{1'})=F_{rr'}(z,z')\exp\left\{i\frac{\pi}{2}(r-r')\right\}
(z')^{\overline Q_{r'}+1}(z)^{-\overline Q_r-\delta_{rr'}}\times\nonumber\\
&&
\exp\left\{\delta_{rr'}\sum_{n>0}\frac{1}{n}\left(\frac{z'}{z}\right)^n\right\},
\label{A.2}
\eey
and
\bey
&& \chi^{\dagger}_{r'}(y^{1'})\chi_r(y^1)=F_{rr'}(z,z')(z')^{\overline
Q_{r'}+\delta_{r',-r}}
(z)^{-\overline Q_r}\times\nonumber\\
&&
\exp\left\{\delta_{rr'}\sum_{n>0}\frac{1}{n}\left(\frac{z}{z'}\right)^n\right\},
\label{A.3}
\eey
with the
$F_{rr'}(z,z')=\frac{1}{2L}e^{-i(\omega_r-\omega_{r'})}e^{i\frac{\pi}{2}
(r-r')\overline Q}:\exp\{2\pi i(r'\partial^{-1}_1( I_{r'}
)_{y^{1'}}-r\partial^{-1}_1( I_r)_{y^1})\}:$. Notice
that $F_{rr}(z,z)=\frac{1}{2L}$.

We see that for $r\not=r'$ the expressions (\ref{A.2}) and (\ref{A.3}) differ
only
by a sign (due to $\exp(i\frac{\pi}{2}(r-r'))=-1$). Hence,
$\{\chi_r(y^1),\chi^{\dagger}_{-r}(y^{1'})\}=0$.
For $r=r'$, we use the analytical
regularization of the type used in \cite{prh}. This yields
\bey
&& \chi_r(y^1)\chi^{\dagger}_r(y^{1'})=\lim_{\varepsilon\to0}\frac{1}{2\pi i}
\oint_{|z^{''}|=1-\varepsilon}\frac{dz^{''}}{z^{''}}\sum_n\left(\frac{z^{''}}{z'}\right)^n
\left(\frac{z^{''}}{z}\right)^{\overline Q_r+1}\times\nonumber\\
&&\times \left(1-\frac{z^{''}}{z}\right)^{-1}F_{rr}(z,z^{''}),\label{A.4}
\eey
and
\bey
&&\chi^{\dagger}_r(y^{1'})\chi_r(y^1)=\lim_{\varepsilon\to0}\frac{1}{2\pi
i}\oint_{|z^{''}|=1
+\varepsilon}\frac{dz^{''}}{z^{''}}\sum_n\left(\frac{z^{''}}{z'}\right)^n
\left(\frac{z^{''}}{z}\right)^{\overline Q_r}\times\nonumber\\
&& \times\left(1-\frac{z}{z^{''}}\right)^{-1}F_{rr}(z,z^{''}).\label{A.5}\eey
Adding (\ref{A.4}) und (\ref{A.5}), we get
\begin{eqnarray}
&&\{\chi_r(y^1),\chi^{\dagger}_r(y^{1'})\}=\lim_{\varepsilon\to0}\frac{1}{2\pi
i}
\left(\oint_{|z^{''}|=1+\varepsilon}-\oint_{|z^{''}|=1-\varepsilon}\right)
\frac{dz^{''}}{z^{''}-z}\times \nonumber\\
&&\times\sum_n\left(\frac{z^{''}}{z'}\right)^n
\left(\frac{z^{''}}{z}\right)^{\overline
Q_r}F_{rr}(z,z^{''})= \nonumber \\
&& \frac{1}{2L}\sum_n\exp\left(\frac{ri\pi
n}{L}(y^1-y^{1'})\right)=\delta(y^1-y^{1'}).
\label{A.6}
\end{eqnarray}
Analogously, one obtains $\{\chi_r(y^1),\chi_{r'}(y^{1'})\}=0$.

To explain eq.~(\ref{42}) let us consider the $\varepsilon$-regularized charge
densities, eq.~(\ref{30}), using eq.~(\ref{A.3}) of the Appendix with the
substitutions: $y^1\to y^1+\frac{ri\varepsilon}{2},y^{1'}\to
y^1-\frac{ri\varepsilon}{2}$,
and expanding in $\varepsilon$ up to $O(\varepsilon^2)$. We get
\bey
I_r(y^1,\varepsilon)&=&\chi^+_r\left(y^1-\frac{ri\varepsilon}{2}\right)\chi_r
\left(y^1+\frac{ri\varepsilon}{2}\right)\exp(r\varepsilon eA_1)\nonumber\\
&=&\frac{1}{2\pi\varepsilon}+I_r(y^1)+\pi\varepsilon(I_r(y^1))^2-\frac{\pi\varepsilon}
{48L^2}+O(\varepsilon^2),\label{A.7}
\eey
in agreement with eq.~(\ref{30}). Differentiating eq.~(\ref{A.7})
with respect to $\varepsilon$, we obtain
\bey
&&\int^L_{-L}dy^1\chi^+_r\left(y^1-\frac{ri\varepsilon}{2}\right)iD_1\chi_r\left(y^1+
\frac{ri\varepsilon}{2}\right)\nonumber\\
&=&-r\left(\frac{L}{\pi\varepsilon^2}+\frac{\pi}{12}\right)+r\pi\int^L_{-L}dy^1
(I_r(y^1))^2+O(\varepsilon),\label{A.8}
\eey
that coincides with eq.~(\ref{42}) after subtracting the constant and taking
the
limit $\varepsilon\to0$.

Eq.~(\ref{44}) is a direct consequence of eq.~(\ref{A.3}) and eq.~(\ref{39}) if
it
is considered on the physical subspace $(Q=0)$. Indeed, from eq.~(\ref{A.3}) we
get
\begin{eqnarray}
&&-\frac{iM}{\eta}(\chi^{\dagger}_+\chi_--\chi^{\dagger}_-\chi_+)=
\frac{iM}{2L\eta}(-1)^Q[e^{i(\omega_+-\omega_-)}e^{i\frac{\pi y^1}{L}Q}
e^{2\pi i(\partial^{-1}_1( I_++I_- ))} \nonumber \\
&&-e^{-i(\omega_+-\omega_-)}e^{-i\frac{\pi y^1}{L}Q}e^{-2\pi i(\partial^{-1}_1
( I_++I_- ))}]  , \label{A.9}
\end{eqnarray}
which indeed coincides with eq.~(\ref{44}) at $Q=0$.

\end{appendix}


\begin{thebibliography}{WWW}
\bibitem{dir} P.A.M. Dirac, Rev. Mod. Phys. {\bf 21}, 392 (1949).
\bibitem{kla} J.R. Klauder, H. Leutwyler and L. Streit, Nuovo Cim. {\bf 66A},
536 (1970).
\bibitem{cas} A. Casher, Phys. Rev. {\bf D14}, 452 (1976).
\bibitem{fra} V.A. Franke, Yu.V. Novozhilov and  E.V. Prokhvatilov, Lett. Math.
Phys.
{\bf 5}, 239, 437 (1981).
\bibitem{per} R.J. Perry, A. Harindranath and K.G. Wilson, Phys. Rev. Lett.
{\bf 65},
2959 (1990).
\bibitem{ann} A.M. Annenkova, E.V. Prokhvatilov and V.A. Franke, Phys. At.
Nucl. {\bf
56} (6), 813 (1993).
\bibitem{ank} A.M. Annenkova, E.V. Prokhvatilov and V.A. Franke, Vestn.
Leningr. Univ.,
Ser. 4, No. 4, 80 (1985) [{\it in Russian}].
\bibitem{bur} M. Burkardt, Phys. Rev {\bf D47}, 4628 (1993).
\bibitem{har} A. Harindranath and J.P. Vary, Phys. Rev. {\bf D36}, 1141 (1987).
\bibitem{wer} T. Heinzl, S. Krusche, S. Simburger and E. Werner, Z. Phys.C {\bf
56}, 415, (1992);
B. van de Sande and S. Pinsky, Phys. Rev. {\bf D49}, 2001 (1994).
\bibitem{byl} A.B. Bylev, E.V. Prokhvatilov and V.A. Franke, Vestn. Leningr.
Univ., Ser.
4, No. 11, 8 (1986) [{\it in Russian}].
\bibitem{ell} T. Eller, H.-C. Pauli and S.J. Brodsky, Phys. Rev. {\bf D35},
1493 (1987).
\bibitem{hor} K. Hornbostel, S.J. Brodsky and  H.-C. Pauli, Phys. Rev. {\bf
D41}, 3814
(1990).
\bibitem{tan} A. Tang, H.-C. Pauli and S.J. Brodsky, Phys. Rev. {\bf D44}, 1842
(1991).
\bibitem{pro} E.V. Prokhvatilov and V.A. Franke, Sov. J. Nucl. Phys. {\bf 49},
688
(1989).
\bibitem{len} F. Lenz, M. Thies, S. Levit and K. Yazaki, Ann. Phys. {\bf 208},
1 (1991).
\bibitem{prk} E.V. Prokhvatilov and V.A. Franke, Sov. J. Nucl. Phys. {\bf 47},
559
(1988).
\bibitem{stev} P.M. Stevenson, Phys. Rev. {\bf D30}, 1712 (1984); Phys. Rev.
{\bf D32}, 1389 (1985).
\bibitem{pol} L. Polley and V. Ritschel, Phys. Lett. {\bf B221}, 44 (1989);
Phys. Rev.
{\bf D45}, No. 8 (1992).
\bibitem{Len1} F. Lenz, H.W.L. Naus, K. Ohta and M. Thies,
{\em Quantum Mechanics of Gauge Fixing},
to be published in  Ann. Phys. (N.Y.).
\bibitem{man} N.S. Manton, Ann. Phys. {\bf 159}, 220 (1985).
\bibitem{prh} E.V. Prokhvatilov, Theor. Math. Phys. {\bf 88}, No. 1 (1991)
({\it
Transl. from Russian}).
\bibitem{byll} A. Bylev, E.V. Prokhvatilov and V.A. Franke, Vestn. Leningr.
Univ., Ser.
4, No. 11, 66 (1989) [{\it in Russian}].
\bibitem{iso} S. Iso and H. Murayama, Prog. Theor. Phys. {\bf 84}, 142 (1990).
\bibitem{grad} I.S. Gradshteyn and I.M. Ryzhik, {\em Tables of Integrals,
Series, and Products}, Academic Press, New York and London, 1965.
\bibitem{fre} I.B. Frenkel and V.N. Kac, Inv. Math. {\bf 62}, 23 (1980).
\bibitem{cole}  S. Coleman, Ann. Phys. (N.Y.) {\bf 101}, 239 (1976).
\bibitem{Len2} F. Lenz, H.W.L. Naus and M. Thies,
{\em QCD in the Axial Gauge Representation},
to be published in  Ann. Phys. (N.Y.).

\end{thebibliography}
\end{document}